\newif\ifDRAFT \DRAFTfalse    
\ifDRAFT
\documentclass[10pt,draftclsnofoot,onecolumn]{IEEEtran}
\else
\documentclass[10pt, conference]{IEEEtran}
\fi
\pdfoutput=1
\usepackage{epsfig}
\usepackage{psfrag}
\usepackage{amssymb, amsmath,mathrsfs}
\usepackage{cite}   
\newcommand{\eref}[1]{(\ref{#1})}                  

\newcommand{\beq}{\begin{equation}}
\newcommand{\eeq}{\end{equation}}

\newcommand{\beqa}{\vspace{0mm} \begin{eqnarray}}
\newcommand{\eeqa}{\end{eqnarray} \vspace{0mm}}

\newcommand{\bi}{\begin{itemize}}
\newcommand{\ei}{\end{itemize}}

\newcommand{\ben}{\begin{enumerate}}
\newcommand{\een}{\end{enumerate}}

\newcommand{\lefto}{\mathopen{}\left}

\newcommand{\sizeparentheses}[1]{\lefto( #1 \right)}
\newcommand{\sizecurly}[1]{\lefto\{ #1 \right\}}

\newcommand{\mt}{\mathrm{M_T}} 					
\newcommand{\mr}{\mathrm{M_R}} 						
\newcommand{\minant}{\mathrm{m}} 			
\newcommand{\maxant}{\mathrm{M}} 				
\newcommand{\channelH}{{\matc{H}}} 
\newcommand{\channelHw}{\matc{H}_{\mspace{-2.0mu}w}} 
\newcommand{\channelHwbar}{\overline{\matc{H}}_{\mspace{-2.0mu}w}}

\newcommand{\setR}{\mathbb{R}}                                  

\newcommand{\cn}[2]{\mathcal{CN}\negthinspace\left(#1,#2\right)}             
\newcommand{\mean}[1]{\mathbb{E}\negthinspace\left\{#1\right\}}             

\newcommand{\mat}[1]{\mathbf{#1}}  
  
\newcommand{\matc}[1]{\bv{\mathcal{#1}}}                             
\newcommand{\bv}[1]{\mbox{\boldmath{$#1$}}}                     
\newcommand{\msf}[1]{\mathsf{#1}}													
\newcommand{\mbb}[1]{\mathbb{#1}}													
\newcommand{\rank}[1]{\mathrm{rank}\negthinspace\left(#1\right)}
\newcommand{\vecop}[1]{\mathrm{vec} \negthinspace\left(#1\right)}             

\newcommand{\expf}[1]{\exp{\left(#1\right)}}                    
\newcommand{\trace}[1]{\mathrm{Tr}\left(#1\right)}              
\newcommand{\diag}[1]{\mathrm{diag}\negthinspace\left\{#1\right\}}             

\newcommand{\mutinf}{\mathrm{I}}                                   
\newcommand{\jensen}{\mathrm{J}}                               
\newcommand{\mutinfexp}[2]{\log\det\negthinspace\left(\mat{I}_{#1}+ #2 \right)}           
  
\newcommand{\snr}{\msf{SNR}}                                 
\newcommand{\dotleq}{\mathrel{\dot{\leq}}} 
\newcommand{\dotgeq}{\mathrel{\dot{\geq}}} 

\newcommand{\prob}[1]{\mathbb{P}\left(#1\right)}
\newcommand{\set}[1]{\mathcal{S}_{#1}}                          
\newcommand{\outage}{\mathcal{O}}  
\newcommand{\joutage}{\mathcal{J}}
\newcommand{\pout}{P_\mathcal{\scriptscriptstyle O}}                               
\newcommand{\pjout}{P_\mathcal{\scriptscriptstyle J}}
\newcommand{\dout}{d_\mathcal{\scriptscriptstyle O}}                               
\newcommand{\djout}{d_\mathcal{\scriptscriptstyle J}}

\newcommand{\codebook}{\mathcal{C}}                              

\newtheorem{tc}{Theorem}

\newtheorem{cc}{Corollary}

\begin{document}
\title{Diversity-Multiplexing Tradeoff in\\ Selective-Fading MIMO Channels}
%
\author{\authorblockN{Pedro Coronel and Helmut B\"olcskei\vspace{.5mm}\\}
\authorblockA{Communication Technology Laboratory\\ 
ETH Zurich, 8092 Zurich, Switzerland \\
E-mail: \{pco, boelcskei\}@nari.ee.ethz.ch}
\thanks{The first author was previously with IBM Research, Zurich Research Laboratory, Switzerland. This work was supported in part by the STREP project No. IST-026905 MASCOT within the Sixth Framework Programme of the European Commission.}
}

\IEEEoverridecommandlockouts

\maketitle
\ifDRAFT
\renewcommand{\baselinestretch}{2.0}
\normalsize
\else
\renewcommand{\baselinestretch}{1}
\normalsize
\fi
\begin{abstract}
We establish the optimal diversity-multiplexing (DM) tradeoff of coherent time, frequency and time-frequency selective-fading MIMO channels and provide a code design criterion for DM-tradeoff optimality. Our results are based on the analysis of the ``Jensen channel'' associated to a given selective-fading MIMO channel. While the original problem seems analytically intractable due to the mutual information being a sum of correlated random variables, the Jensen channel is equivalent to the original channel in the sense of the DM-tradeoff and lends itself nicely to analytical treatment. Finally, as a consequence of our results, we find that the classical rank criterion for space-time code design (in selective-fading MIMO channels) ensures optimality in the sense of the DM-tradeoff. 
\end{abstract}
\section{Introduction}

The diversity-multiplexing (DM) tradeoff framework introduced by Zheng and Tse \cite{ZheTse02} allows to efficiently characterize the information-theoretic performance limits of communication over multiple-input multiple-output (MIMO) fading channels. In addition, the results in \cite{ZheTse02} have triggered significant activity on the design of DM-tradeoff optimal space-time codes. In particular, the \textit{non-vanishing} determinant criterion \cite{BelRek03, YaoWor03} on codeword difference matrices has been shown to constitute a sufficient condition for DM-tradeoff optimality in flat-fading MIMO channels with two transmit and two or more receive antennas \cite{YaoWor03}; this criterion has led to the construction of space-time codes based on constellation rotation \cite{YaoWor03, DayVar05} and cyclic division algebras \cite{BelRekVit04}. In \cite{GamCaiDam04} lattice-based space-time codes have been shown to be DM-tradeoff optimal. The DM-tradeoff optimality of \textit{approximately universal} space-time codes was established in \cite{TW05}.  

\textit{Contributions:}
While the results mentioned above focus on frequency-flat block-fading channels, extensions to frequency-selective channels can be found in \cite{GrokopTse04, MedSlo05}. However, a general characterization of the optimal DM-tradeoff in time, frequency or time-frequency selective-fading MIMO channels, in the following simply referred to as selective-fading MIMO channels, remains an open problem. The present paper resolves this problem for the coherent case (i.e., for perfect channel state information (CSI) at the receiver) and provides a code design criterion guaranteeing DM-tradeoff optimality. Our results are based on exponentially tight (in the sense of exhibiting the same DM-tradeoff behavior) upper and lower bounds on the mutual information of (coherent) selective-fading MIMO channels. In particular, we show that the DM-tradeoff of this class of channels can be obtained by solving the analytically tractable problem of computing the DM-tradeoff curve corresponding to the associated ``Jensen channel''.

\textit{Notation:} $\mt$ and $\mr$ denote the number of transmit and receive antennas, respectively. We define $\minant\negmedspace:=\min(\mt,\mr)$ and $\maxant\negmedspace:= \negmedspace\max(\mt,\mr)$. For $x\negmedspace\in\negmedspace \mathbb{R}$, we let $[x]^+ \negmedspace := \negmedspace\max{(0, x)}$. The superscripts ${}^T$, ${}^H$ and ${}^*$ stand for transposition, conjugate transposition and complex conjugation, respectively. $\mat{I}_n$ is the $n\times n$ identity matrix, $\mat{A}\otimes \mat{B}$ and $\mat{A}\odot \mat{B}$ denote, respectively, the Kronecker and Hadamard products of the matrices $\mat{A}$ and $\mat{B}$, and $\mat{A}\succeq\mat{B}$ stands for the positive semidefinite ordering. If $\mat{A}$ has columns $\mat{a}_k$ ($k \negmedspace = \negmedspace 1, 2, \ldots, m$), $\vecop{\mat{A}}=[\mat{a}_1^T \:  \mat{a}_2^T \: \ldots \: \mat{a}_m^T]^T$. For the $n\times m$ matrices $\mat{A}_k$ ($k\negmedspace=\negmedspace0, 1, \ldots, K-1$), $\diag{\mat{A}_k}_{k=0}^{K-1}$ denotes the $nK\times mK$ block-diagonal matrix with the $k$th diagonal entry given by $\mat{A}_k$. If $\set{}$ is a set, $|\set{}|$ denotes its cardinality. For index sets $\set{1} \subseteq \sizecurly{1, 2, \ldots, n}$ and $\set{2} \subseteq \sizecurly{1, 2, \ldots, m}$, $\mat{A}(\set{1},\set{2})$ stands for the (sub)matrix consisting of the rows of $\mat{A}$ indexed by $\set{1}$ and the columns of $\mat{A}$ indexed by $\set{2}$. The eigenvalues of the $n\times n$ Hermitian matrix $\mat{A}$, sorted in ascending order, are denoted by $\lambda_k(\mat{A})$, $k \negmedspace = \negmedspace 1, 2, \ldots, n$. The Kronecker delta function is defined as $\delta(m)\negmedspace=\negmedspace1$ for $m=0$ and zero otherwise. If  $X$ and $Y$ are random variables (RVs), $X\sim Y$ denotes equality in distribution and $\mathbb{E}_X$ is the expectation operator with respect to (w.r.t.) the RV $X$. The random vector $\mat{x} \sim \cn{\mat{0}}{\mat{C}}$ is multivariate circularly symmetric zero-mean complex Gaussian with $\mean{\mat{x}\mat{x}^H}=\mat{C}$. $f(x)$ and $g(x)$ are said to be exponentially equal, denoted by $f(x)\doteq g(x)$, if $\lim_{x\rightarrow \infty} \frac{\log f(x)}{\log x} = \lim_{x\rightarrow \infty} \frac{\log g(x)}{\log x}$. Exponential inequality, denoted by $\:\dotgeq$ and $\dotleq$, is defined analogously. 

\section{Channel and signal model\label{Sec.Model}}

The input-output relation for the class of MIMO channels considered in this paper is given by 
\beqa
\mat{y}_n = \sqrt{\frac{\snr}{\mt}} \mat{H}_n \mat{x}_{n} +\mat{z}_n,\quad n=0, 1, \ldots, N-1\label{Eq.SigModel}
\eeqa
where the index $n$ corresponds to a time, frequency or time-frequency slot and SNR denotes the signal-to-noise ratio at each receive antenna. The vectors $\mat{y}_n $, $\mat{x}_n$ and  $\mat{z}_n$ denote, respectively, the corresponding $\mr \times 1$ receive signal vector, $\mt \times 1$ transmit signal vector, and $\mr \times 1$ zero-mean circularly symmetric complex Gaussian noise vector satisfying $\mean{\mat{z}_n\mat{z}_{n' }^H}= \delta(n-n')\: \mat{I}_{\mr}$. We restrict our analysis to spatially uncorrelated Rayleigh fading channels so that, for a given $n$, $\mat{H}_n$ has i.i.d. $\cn{0}{1}$ entries. We do allow, however, for correlation across $n$, assuming, for simplicity, that each scalar subchannel has the same correlation function, i.e., $\mean{\mat{H}_n(i,j) (\mat{H}_{n-m}(i,j))^*}=r_{\mbb{H}}(m)$, $(i\negmedspace=\negmedspace1, 2, \ldots, \mr,  j\negmedspace=\negmedspace1, 2, \ldots, \mt)$. Defining $\mat{H} = [\mat{H}_0 \: \mat{H}_1 \:  \ldots \: \mat{H}_{N-1}]$, we therefore have
\beqa
\mean{\vecop{\mat{H}}(\vecop{\mat{H}})^H} = \mat{R}_{\mbb{H}} \otimes\mat{I}_{\mt\mr}\label{Eq.CovModel}
\eeqa
where the covariance matrix $\mat{R}_\mathbb{H}(i,j)\negmedspace=\negmedspace r_\mbb{H}(i\negmedspace-\negmedspace j)$ $(i,j\negmedspace=\negmedspace0,1, \ldots, N\negmedspace-\negmedspace 1)$ follows from the channel's scattering function \cite{Bello63}. In the purely frequency-selective case, e.g., assuming an orthogonal frequency-division multiplexing (OFDM) system \cite{PelRui80} with $N$ tones and hence $\mat{H}_n = \sum_{l=0}^{L-1} \mat{H}(l) \:e^{-j\frac{2 \pi}{N}{l n}}$, where the uncorrelated (across $l$) matrix-valued taps $\mat{H}(l)$ have i.i.d. $\cn{0}{\sigma_l^2}$ entries, we obtain $r_\mbb{H}(m) = \sum_{l=0}^{L-1} \sigma_l^2 e^{-j \frac{2 \pi}{N} l m}$ $(m\negmedspace=\negmedspace0, 1, \ldots, N\negmedspace-\negmedspace1)$. In the remainder of the paper, we use the definition $\rho\negthinspace:=\negthinspace\rank{\mat{R}_\mbb{H}}$.

\section{Diversity-multiplexing tradeoff\label{Sec.DMtradeoff}}

\subsection{Preliminaries}
Assuming perfect CSI in the receiver, the mutual information of the channel in \eref{Eq.SigModel} is given by
\begin{equation}
\mutinf \sizeparentheses{\snr} = \frac{1}{N}\sum_{n=0}^{N-1} \mutinfexp{\mr}{\frac{\snr}{\mt} \mat{H}_{n} \mat{C}_{n} \mat{H}_{n}^H}\label{Eq.Mutinf}
\end{equation}
where the transmit signal vectors are uncorrelated across $n$ and satisfy $\mat{x}_{n} \sim \cn{\mat{0}}{\mat{C}_{n}}$ with power constraint $\trace{\mat{C}_{n}}\leq \mt$, $n\negmedspace=\negmedspace0, 1, \ldots, N\negmedspace-\negmedspace1$. The DM-tradeoff realized by a family (w.r.t. SNR) of codes $\codebook_r$ with rate $R(\snr)=r \log\snr$, where $r\in [0,\minant]$, is given by the function 
\begin{equation*}
d_{\codebook}(r)=-\lim_{\snr\rightarrow\infty}\frac{\log P_e(r,\snr)}{\log\snr}
\end{equation*} 
where $P_e(r,\snr)$ is the error probability obtained through ML detection. At a given SNR, the corresponding codebook $\codebook_r(\snr)$ contains $\snr^{Nr}$ codewords $\mat{X}=[\mat{x}_0\; \mat{x}_1\; \ldots \; \mat{x}_{N-1}]$. We say that such a family of codes $\codebook_r$ operates at multiplexing rate $r$. The optimal tradeoff curve $d^\star\mspace{-2.0mu}(r) = \sup_{\codebook_r} d_{\codebook}(r)$, where the supremum is taken over all families of codes satisfying $R(\snr)= r \log\snr$, quantifies the maximum achievable diversity gain as a function of $r$. Since the outage probability $\pout(r,\snr)$ is a lower bound to the error probability \cite{ZheTse02}, we have
\begin{equation*}
d^\star\mspace{-2.0mu}(r) \leq \dout(r) =  - \lim_{\snr\rightarrow\infty}\frac{\log\pout(r, \snr)}{\log\snr}.
\end{equation*}
Extending the arguments that lead to \cite[Eq. (9)]{ZheTse02} to the case $N\negmedspace>\negmedspace1$, we can conclude that setting $\mat{C}_{n}\negmedspace =\negmedspace\mat{I}_{\mt}$ ($n=0, 1, \ldots, N\negmedspace-\negmedspace1$) in \eref{Eq.Mutinf} does not alter the exponential behavior of mutual information. Hence
\begin{align}
\pout&(r, \snr) \doteq\notag \\ &\prob{\frac{1}{N}\sum_{n=0}^{N-1} \mutinfexp{\mr}{\snr \:\mat{H}_{n}\mat{H}_{n}^H}<  r \log\snr}\label{Eq.OutProb}
\end{align}
where we used the fact that the factor $1/\mt$ in \eref{Eq.Mutinf} can be neglected in the scale of interest. Let $\bv{\mu}(n):=[\mu_1(n)\: \mu_2(n)\: \ldots\: \mu_{\minant}(n)]$ ($n=0,1,\ldots,N-1$), with the singularity levels defined as
\begin{equation*}
\mu_k(n) = -\frac{\log\lambda_k(\mat{H}_n \mat{H}_n^H)}{\log\snr}, \quad k=1, 2, \ldots, \minant
\end{equation*}
and note that \cite{ZheTse02}
\begin{equation}
\pout(r, \snr) \doteq \prob{\outage(r)}
\end{equation}
where
\begin{multline}
\outage(r)=\bigg\{ \bv{\mu}(n)\in \mathbb{R}_+^{\minant}, n=0,1,\ldots,N-1 :\\ \frac{1}{N}\sum_{n=0}^{N-1}\sum_{k=1}^{\minant} \: [1-\mu_k(n)]^+ \negthinspace<  r\bigg\}\label{Eq.OutEvent}
\end{multline}
and $\mathbb{R}_+^{\minant}$ denotes the nonnegative orthant. Unlike the frequency-flat fading case treated in \cite{ZheTse02}, characterizing $\dout(r)$ for the selective-fading case seems analytically intractable with the main difficulty stemming from the fact that one has to deal with the sum of correlated (recall that the $\mat{H}_n$ are correlated across $n$) terms in \eref{Eq.OutProb}. It turns out, however, that one can find lower and upper bounds on $\mutinf(\snr)$ which are exponentially tight (and, hence, preserve the DM-tradeoff behavior) and analytically tractable. The next section formalizes this idea.

\subsection{Jensen channel and Jensen outage event}\label{Subsection.Jout}

We start by noting that applying Jensen's inequality yields
\begin{multline}
\mutinf(\snr)=\frac{1}{N} \sum_{n=0}^{N-1} \mutinfexp{\mr}{\frac{\snr}{\mt}\mat{H}_n\mat{H}_n^H}
\leq \\ \log\det\sizeparentheses{\mat{I}_{\minant}+\frac{\snr}{\mt N}\channelH\channelH^H} := \jensen(\snr) \label{Eq.Jensen}
\end{multline}
where the ``Jensen channel" is defined as
\begin{equation*}
\channelH = \begin{cases} \; [\mat{H}_0 \; \mat{H}_1\; \ldots \; \mat{H}_{N-1}], & \text{if $\mr\leq\mt$,} \\ [\mat{H}_0^H \; \mat{H}_1^H\; \ldots \; \mat{H}_{N-1}^H], & \text{if $\mr > \mt$.}\end{cases} \label{Eq.EqChannel}
\end{equation*}
In the following, we say that a Jensen outage event occurs if the Jensen channel $\channelH$ is in outage w.r.t. the rate $R(\snr)=r\log\snr$, i.e., if $\jensen(\snr)<R(\snr)$. The corresponding outage probability will be denoted as $\pjout(r, \snr)$ and clearly satisfies $\pjout(r,\snr)\leq\pout(r,\snr)$. The operational significance of a Jensen outage will be established at the end of this section. We shall first focus on characterizing the Jensen outage event analytically. Using \eref{Eq.CovModel}, it is readily seen that $\channelH= \channelHw(\mat{R}_{\mbb{H}}^{1/2} \otimes \mat{I}_{\maxant})$, where $\channelHw$ is an i.i.d. $\cn{0}{1}$ matrix with the same dimensions as $\channelH$. Noting that $\channelHw \mat{U} \sim \channelHw$ for $\mat{U}$ unitary and using the eigendecomposition $\mat{R}_{\mbb{H}}\otimes \mat{I}_{\maxant} = \mat{U} (\mat{\Lambda}\otimes \mat{I}_{\maxant}) \mat{U}^H$, where $\mat{\Lambda}=\diag{\lambda_1(\mat{R}_\mbb{H}), \lambda_2(\mat{R}_\mbb{H}), \ldots, \lambda_\rho(\mat{R}_\mbb{H}),0,\ldots, 0}$, it follows that
\begin{align*}
\jensen(\snr)& = \mutinfexp{{\minant}}{\frac{\snr}{\mt N}\channelHw (\mat{R}_{\mbb{H}}\otimes \mat{I}_{\maxant})\channelHw^H}\\
&\sim \mutinfexp{{\minant}}{\frac{\snr}{\mt N} \channelHw \sizeparentheses{\mat{\Lambda}\otimes\mat{I}_{\maxant}}\channelHw^H}.
\end{align*}
Next, observe that the following positive semidefinite ordering holds
\begin{equation}
\lambda_1(\mat{R}_\mbb{H})\:\diag{\mat{I}_{\rho\maxant}, \mat{0}} \; \preceq\;
\mat{\Lambda}\otimes\mat{I}_{\maxant}\;  \preceq \lambda_\rho(\mat{R}_\mbb{H})\: \diag{\mat{I}_{\rho\maxant}, \mat{0}}.\label{Eq.ordering}
\end{equation}
Since $f(\mat{A})= \log\det(\mat{I}+\mat{A})$ is increasing over the cone of positive semidefinite matrices \cite{BV}, we get the following bounds on the Jensen outage probability
\begin{align} \label{Eq.ULB}\begin{split}
&\prob{\mutinfexp{\minant}{\lambda_\rho(\mat{R}_\mbb{H})\frac{\snr}{\mt N}\channelHwbar \channelHwbar^H}< r \log\snr} \\
&\leq \pjout(r, \snr)\\
&\leq  \prob{\mutinfexp{\minant}{\lambda_1(\mat{R}_\mbb{H})\frac{\snr}{\mt N}\channelHwbar\channelHwbar^H}< r \log\snr}
\end{split}\end{align}
where $\channelHwbar = \channelHw([1\negmedspace:\negmedspace\minant], [1\negmedspace:\negmedspace\rho\maxant])$. Taking the exponential limit (in SNR) in \eref{Eq.ULB}, it follows readily that
\begin{equation}
\pjout(r,\snr) \doteq \prob{\log\det\sizeparentheses{\mat{I}_{\minant}+ \snr \:\channelHwbar\channelHwbar^H}\negmedspace< r \log\snr}\negthickspace.\label{Eq.Outprob2}
\end{equation}
For later use, we define $\bv{\alpha} :=[\alpha_1\:\alpha_2\:\ldots\:\alpha_\minant]$ with the singularity levels
\begin{equation}
\alpha_k\negmedspace=\negmedspace-\frac{\log \lambda_k(\channelHwbar\channelHwbar^H)}{\log\snr}, \quad k=1, 2, \ldots, \minant \label{Eq.alpha}
\end{equation}
and note that $\pjout(r, \snr) \doteq \prob{\joutage(r)}$, where
\begin{gather*}
\joutage(r) = \sizecurly{\bv{\alpha}\in \setR_+^{\minant}\negthickspace: \alpha_1\negthinspace \geq\negthinspace \alpha_2\negthinspace \geq\negthinspace \ldots \negthinspace \geq \negthinspace \alpha_{\minant}, \sum_{k=1}^{\minant} [1-\alpha_k]^+ < r}.
\end{gather*}
It is now natural to define the Jensen outage curve as
\beqa
\djout(r) = - \lim_{\snr\rightarrow\infty}\frac{\log\pjout(r, \snr)}{\log\snr}.\nonumber
\eeqa
Based on \eref{Eq.Outprob2}, we can conclude that $\djout(r)$ is nothing but the DM-tradeoff curve of an effective MIMO channel with $\rho\maxant$ transmit and $\minant$ receive antennas. We can therefore directly apply the results in \cite{ZheTse02} to infer that the Jensen outage curve is the piecewise linear function connecting the points $(r,\djout(r))$ for $r=0, 1, \ldots, \minant$, with
\begin{equation}
\djout(r) = (\rho\maxant-r)(\minant-r).\label{Eq.JensenCurve}
\end{equation}
Since, as already noted, $\pjout(r,\snr) \leq \pout(r, \snr)$, we obtain
\begin{equation}
d_{\codebook}(r) \leq d^\star\mspace{-2.0mu}(r) \leq \dout(r) \leq \djout(r), \quad r\in [0, \minant],\label{Eq.DivOrders}
\end{equation}
for any family of codes $\codebook_r$. The optimal DM-tradeoff curve $d^\star\mspace{-2.0mu}(r)$ will be established in the next section by showing that codes satisfying $d_{\codebook}(r) =\djout(r)$ do exist and hence $d^\star\mspace{-2.0mu}(r)=\djout(r)$.

\section{Jensen-optimal code design criterion\label{Sec.CDC}}

The goal of this section is to derive a sufficient condition for a family of codes to achieve $\djout(r)$, and hence, by virtue of \eref{Eq.DivOrders}, to be DM-tradeoff optimal.

\subsection{Code design criterion}
\begin{tc}\label{Th.CDC}
Consider a family of codes $\codebook_r$ with block length $N\geq\rho\mt$ that operates over the channel \eref{Eq.SigModel}. If, for any codebook $\codebook_r(\snr) \in \codebook_r$ and any two codewords $\mat{X}$, $\mat{X}' \in \codebook_r(\snr)$, the codeword difference matrix $\mat{E}=\mat{X}-\mat{X}'$ is such that 
\begin{equation}
\rank{\mat{R}_{\mbb{H}} \odot \mat{E}^H\mat{E}}= \rho\mt\label{Eq.ThCDC}
\end{equation}
then the error probability (for ML decoding) satisfies
\begin{equation*}
P_e(r, \snr)\doteq \snr^{-\djout(r)}.
\end{equation*}\end{tc}
\vspace{3mm}
\begin{proof} 
We start by deriving an upper bound on the average (w.r.t. the random channel) pairwise error probability (PEP). Assuming that $\mat{X}$ was transmitted, the probability of the ML decoder mistakenly deciding in favor of codeword $\mat{X}'$ can be upper-bounded in terms of the codeword difference vectors $\mat{e}_n=\mat{x}_n-\mat{x}_n'$ ($n=0,1,\ldots,N-1$) as
\begin{align*}
\prob{\mat{X}\rightarrow\mat{X}'} &\leq \mathbb{E}_{\mat{H}}\sizecurly{\expf{-\frac{\snr}{4\mt}\sum_{n=0}^{N-1}||\mat{H}_n \mat{e}_n||^2}}\\
&=\mathbb{E}_{\mat{H}}\sizecurly{\expf{-\frac{\snr}{4\mt}\trace{\mat{H}_w\mat{\Upsilon}\mat{H}_w^H}}}
\end{align*}
where
\begin{equation*}
\mat{\Upsilon}=(\mat{R}_{\mbb{H}}^{1/2} \otimes \mat{I}_{\mt}) \:\diag{\mat{e}_n\mat{e}_n^H}_{n=0}^{N-1}\: \negthinspace(\mat{R}_{\mbb{H}}^{1/2} \otimes \mat{I}_{\mt})
\end{equation*}
and $\mat{H}_w$ denotes an $\mr \times \mt N$  i.i.d. $\cn{0}{1}$ matrix. Straightforward manipulations reveal that $\rank{\mat{\Upsilon}}\negmedspace=\negmedspace\rank{\mat{R}_{\mbb{H}} \odot \mat{E}^H\mat{E}}$ so that the assumption \eref{Eq.ThCDC} implies $\rank{\mat{\Upsilon}} \negmedspace=\negmedspace\rho\mt$. With the eigendecomposition $\mat{\Upsilon} \negmedspace = \negmedspace\mat{U\Lambda U}^H$, we have $\trace{\mat{H}_w \mat{\Upsilon}\mat{H}_w^H} \negmedspace \sim \negmedspace\trace{\mat{H}_w \mat{\Lambda}\mat{H}_w^H}$, and hence
\begin{equation*}
\prob{\mat{X}\rightarrow\mat{X}'} \nonumber\\\leq \mathbb{E}_{\mat{H}}\sizecurly{\expf{-\frac{\snr}{4\mt}\trace{\mat{H}_w \mat{\Lambda}\mat{H}_w^H}}}.
\end{equation*}
Setting $\overline{\mat{H}}_w = \mat{H}_w([1\negthickspace:\negthickspace\mr],[1\negthickspace:\negthickspace\rho\mt])$ and denoting the smallest nonzero eigenvalue of $\mat{\Upsilon}$ as $\lambda$, we note that
\begin{equation}
\trace{\mat{H}_w\mat{\Lambda}\mat{H}_w^H} \geq \lambda  \; \trace{\overline{\mat{H}}_w \overline{\mat{H}}_w^H}\label{Eq.IneqProof}
\end{equation}
and thus
\begin{equation}
\prob{\mat{X}\rightarrow\mat{X}'} \leq  \mathbb{E}_{\overline{\mat{H}}_w}\sizecurly{\expf{- \frac{\lambda\:\snr\:}{4\mt} \trace{\overline{\mat{H}}_w \overline{\mat{H}}_w^H}}}.\label{Eq.y}
\end{equation}
Next, note that
\begin{align}
\trace{\overline{\mat{H}}_w \overline{\mat{H}}_w^H}  \; &=\; \trace{\channelHwbar \channelHwbar^H}\notag\\
&= \; \sum_{k=1}^{\minant} \lambda_k(\channelHwbar \channelHwbar^H)\notag\\
&= \; \sum_{k=1}^{\minant} \snr^{-\alpha_k} \label{Eq.x}
\end{align}
where \eref{Eq.x} follows from \eref{Eq.alpha}. We can now write the PEP upper-bound in \eref{Eq.y} in terms of the singularity levels $\alpha_k$ ($k\negmedspace=\negmedspace1, 2, \ldots, \minant$) characterizing the Jensen outage event:
\begin{equation}
\prob{\mat{X}\rightarrow\mat{X}'} \leq  \mathbb{E}_{\bv{\scriptstyle{{\alpha}}}}\sizecurly{\expf{-\frac{\lambda}{4\mt}\sum_{k=1}^{\minant} \snr^{1-\alpha_k}}}.\label{Eq.PEPbound1}
\end{equation}
Next, consider a realization of the random vector $\bv{\alpha}$ and let $\set{}=\{k: \alpha_k\leq1\}$. We have
\begin{align}
\sum_{k=1}^{\minant} \snr^{1-\alpha_k} &\geq \sum_{k\in \set{}} \snr^{1-\alpha_k}\notag\\
&\begin{subarray}{c} \mathrm{\scriptscriptstyle{(i)}} \\\geq \end{subarray} \quad |\set{}| \; \snr^{\frac{1}{|\set{}|} \sum_{k\in \set{}}{(1-\alpha_k)}}\notag\\
&\begin{subarray}{c} \mathrm{\scriptscriptstyle{(ii)}} \\ = \end{subarray} \quad |\set{}| \; \snr^{\frac{1}{|\set{}|} \sum_{k=1}^{\minant}{[1-\alpha_k]^+}} \label{Eq.localpha}
\end{align}
where $\mathrm{(i)}$ follows from the arithmetic-geometric mean inequality and $\mathrm{(ii)}$ follows from the definition of $\set{}$. Using \eref{Eq.localpha} in \eref{Eq.PEPbound1}, we get
\begin{equation}
\prob{\mat{X}\rightarrow\mat{X}'} \leq   \mathbb{E}_{\bv{\scriptstyle{{\alpha}}}}\sizecurly{\expf{-\frac{\lambda\: |\set{}|}{4\mt}\:  \snr^{\frac{1}{|\set{}|}\sum_{k=1}^{\minant}[1-\alpha_k]^+}}}.\label{Eq.PEPbound}
\end{equation}
The dependence of the PEP upper bound \eref{Eq.PEPbound} on the singularity levels characterizing the Jensen outage event suggests to split up the overall error probability according to
\begin{align}
P_e(r, \snr) &= \prob{\mathrm{error}, \bv{\alpha} \in \joutage(r)} + \prob{\mathrm{error}, \bv{\alpha} \notin \joutage(r)}\notag\\
&= \prob{\bv{\alpha} \in \joutage(r)} \prob{\mathrm{error}| \bv{\alpha} \in \joutage(r)} \notag\\
&{\hspace{7mm}}+ \prob{\bv{\alpha} \notin \joutage(r)} \prob{\mathrm{error}| \bv{\alpha} \notin \joutage(r)}\notag\\
&\leq  \prob{\bv{\alpha} \in \joutage(r)}\notag\\ &{\hspace{7mm}}+ \prob{\bv{\alpha} \notin \joutage(r)}\prob{\mathrm{error} | \bv{\alpha}\notin \joutage(r)} 
.\label{Eq.UpBoundErrorProb}
\end{align}
For any $\bv{\alpha} \notin \joutage(r)$, we have $\sum_{k=1}^{\minant}[1-\alpha_k]^+ \geq r$ and $|\set{}| \geq 1$, which upon noting that $|\codebook_r(\snr)|=\snr^{Nr}$, yields the following union bound based on the PEP in \eref{Eq.PEPbound}
\begin{align*}
\prob{\mathrm{error}|\bv{\alpha} \notin \joutage(r)} &\leq \snr^{Nr} \expf{- \frac{\lambda}{4\mt}  \: \snr^{r/\minant}}
\end{align*}
where we used $|\set{}| \leq \minant$. Hence, for any $r>0$, $\prob{\mathrm{error}|\bv{\alpha} \notin \joutage(r)}$ decays exponentially in SNR and we have
\begin{align}\label{Eq.UnionBound}
\prob{\mathrm{error}, \bv{\alpha} \notin \joutage(r)} &=\underbrace{\prob{\bv{\alpha} \notin \joutage(r)} }_{\leq 1}\prob{\mathrm{error}| \bv{\alpha} \notin \joutage(r)}\notag\\
&\leq \snr^{Nr}\expf{- \frac{\lambda}{4\mt}\:\snr^{r/\minant}}.
\end{align}
Consequently, noting that $\prob{\bv{\alpha}\in \joutage(r)} \doteq \pjout(r, \snr)$ and using \eref{Eq.UnionBound} in \eref{Eq.UpBoundErrorProb}, we obtain
\begin{equation*}
P_e(r, \snr) \dotleq \pjout(r, \snr).
\end{equation*}
Since $\pjout(r, \snr) \leq \pout(r, \snr)$, it follows trivially that $\pjout(r,\snr) \dotleq \pout(r, \snr)$. In addition, for a specific family of codes $\codebook_r$, we have $\pout(r,\snr)\leq P_e(r, \snr)$ and hence $\pout(r, \snr)\dotleq P_e(r,\snr)$. Putting the pieces together, we finally obtain
\begin{equation*}
\pout(r,\snr)\dotleq P_e(r,\snr)\dotleq \pjout(r,\snr)\dotleq \pout(r,\snr)
\end{equation*}
which implies
\begin{equation*}
P_e(r,\snr)\doteq \pjout(r, \snr)
\end{equation*}
and hence (by definition of $\djout(r)$)
\begin{equation*}
P_e(r,\snr)\doteq \snr^{-\djout(r)}.
\end{equation*}
\end{proof}

As a direct consequence of Theorem \ref{Th.CDC}, a family of codes that satisfies \eref{Eq.ThCDC} for all codeword difference matrices in any codebook $\codebook_r(\snr) \in \codebook_r$ realizes a DM-tradeoff curve $d_\codebook(r)=\djout(r)$ and hence, by \eref{Eq.DivOrders}
\begin{equation*}
\djout(r) \leq d^\star\mspace{-2.0mu}(r) \leq \djout(r)
\end{equation*}
which implies
\begin{equation}
d^\star\mspace{-2.0mu}(r) = \djout(r).\label{Eq.CurvesEqual}
\end{equation}
The optimal DM-tradeoff curve for selective-fading MIMO channels is therefore given by the DM-tradeoff curve of the associated Jensen channel. Put differently, Theorem \ref{Th.CDC} shows that, even though $\joutage(r) \subseteq \outage(r)$ by definition, we still have
\begin{equation*}
\prob{\joutage(r)} \doteq \prob{\outage(r)}
\end{equation*}
which essentially says that the ``original'' channel has the same high-SNR outage behavior as its associated Jensen channel. 

The code design criterion in Theorem \ref{Th.CDC} provides a sufficient condition for achieving the DM-tradeoff curve. Interestingly, the classical rank criterion \cite{TarSesCal98, Tar99, Bol00, BolISIT02, BolBorPau03,MaLeuGia05}, aimed at maximizing the diversity gain for $r=0$, can be shown \cite{pcoj06} to be equivalent to the criterion in Theorem \ref{Th.CDC}. We emphasize, however, that optimality w.r.t. the DM-tradeoff at multiplexing rate $r$ requires that \eref{Eq.ThCDC} is satisfied for all codeword difference matrices in any codebook $\codebook_r(\snr) \in \codebook_r$, in particular also for $\snr\rightarrow\infty$. We next state a sufficient condition for DM-tradeoff optimality which makes this aspect explicit and establishes a connection to the approximately universal code design criterion in \cite{TW05}.

\begin{cc} \label{Cor1}
A family of codes $\codebook_r$ of block length $N\geq \rho\mt$ is DM-tradeoff optimal if there exists an $\epsilon>0$ such that
\begin{equation}\label{Eq.DecayRate}
\lambda^\minant(\snr) \dotgeq \snr^{-(r-\epsilon)}
\end{equation}
where
\begin{equation*}
\lambda(\snr)=\min_{\begin{subarray}{c} k=1, 2,\ldots,\rho\mt \\ \mat{E}=\mat{X}-\mat{X}',\:\mat{X},\mat{X}'\in \codebook_r(\snr)\end{subarray}} \bigg\{\lambda_k(\mat{R}_{\mbb{H}} \odot \mat{E}^H\mat{E})>0\bigg\}.
\end{equation*}
\end{cc}
\vspace{2mm}
\begin{proof}
Using \eref{Eq.DecayRate} in \eref{Eq.UnionBound}, we obtain
\begin{equation*}
\prob{\mathrm{error}, \bv{\alpha} \notin \joutage(r)}\leq \snr^{Nr}\expf{- \frac{\snr^{\epsilon/\minant}}{4\mt}}
\end{equation*}
which, following the same logic as in the proof of Theorem \ref{Th.CDC}, implies that $P_e(r,\snr) \doteq \snr^{-\djout(r)}$. 
\end{proof}

Note that the quantity $\lambda^\minant(\snr)$ is trivially a lower bound on the product of the $\minant$ smallest nonzero eigenvalues of any codeword difference matrix in the codebook $\codebook_r(\snr)$. Consequently, in the case of non-selective fading, where $\mat{R}_\mbb{H} \odot \mat{E}^H\mat{E}= \mat{E}^H\mat{E}$, any family of codes $\codebook_r$ satisfying \eref{Eq.DecayRate} will also be approximately universal in the sense of \cite[Th. 3.1]{TW05}. Moreover, if $\lambda(\snr)$ remains strictly positive as $\snr\rightarrow\infty$, $\codebook_r$ fulfills the non-vanishing determinant criterion \cite{BelRek03,YaoWor03} and will, by \eref{Eq.UnionBound} and the same arguments as in the proof of Theorem \ref{Th.CDC}, be DM-tradeoff optimal. 
\subsection{Application to the frequency-selective case}
As an example, we shall next specialize our results to frequency-selective fading MIMO channels, recovering the results reported previously in \cite{GrokopTse04,MedSlo05}. For the sake of simplicity of exposition, we shall employ a cyclic signal model, as obtained in an OFDM system for example. The channel's transfer function is given by
\begin{equation*}
\mat{H}(e^{j 2\pi \theta}) = \sum_{l=0}^{L-1} \mat{H}(l)\:  e^{-j 2\pi l \theta}, \quad 0\leq \theta < 1
\end{equation*}
where the $\mat{H}(l)$ have i.i.d. $\cn{0}{\sigma_l^2}$ entries and satisfy
\begin{equation*}
\mean{\vecop{\mat{H}(l)}\vecop{\mat{H}(l')}^H}=\sigma_l^2 \: \delta(l-l')\:\mat{I}_{\mt\mr}.
\end{equation*}
With $\mat{H}_n = \mat{H}(e^{j2\pi\frac{n}{N}})$, $n\negmedspace=\negmedspace0, 1 \ldots, N-1$, the channel's covariance matrix follows as
\begin{equation*}
\mat{R}_\mbb{H} = \mat{F}\:\diag{\sigma_0^2, \sigma_1^2, \ldots, \sigma_{L-1}^2, 0, \ldots, 0}\mat{F}^H
\end{equation*}
where $\mat{F}$ is the $N\times N$ FFT matrix. Since $\rank{\mat{R}_\mbb{H}}=L$, inserting $\rho=L$ into \eref{Eq.JensenCurve} and using \eref{Eq.CurvesEqual} yields the optimal DM-tradeoff curve as the piecewise linear function connecting the points $(r, d^\star\mspace{-2.0mu}(r))$ for $ r=0,1, \ldots, \minant$, with
\begin{equation}
d^\star\mspace{-2.0mu}(r) =(L\maxant - r) (\minant-r).\label{Eqloc}
\end{equation}
This is the optimal DM-tradeoff curve for frequency-selective fading MIMO channels reported previously in \cite{MedSlo05}. Specializing \eref{Eqloc} to the case $\mt\negthinspace=\negthinspace \mr\negthinspace=\negthinspace 1$ and noting that $d^\star\mspace{-2.0mu}(r)=(L-r)(1-r)=L(1-r)$ for $r=\{0, 1\}$, yields the results reported in \cite{GrokopTse04}. We note that the proof techniques employed in \cite{GrokopTse04,MedSlo05} are different from the approach taken in this paper and seem to be tailored to the frequency-selective case. In addition, our approach is not limited to large code lengths as \eref{Eq.ThCDC} can be guaranteed for any $N\geq L\mt$.


\section{Conclusions\label{Sec.Conclusion}}
\enlargethispage{-1.03in}
Analyzing the high-SNR outage behavior of the Jensen channel instead of the original channel was found to be an effective tool to establish the DM-tradeoff in selective-fading MIMO channels. We showed that satisfying extensions (to the selective-fading MIMO case) of the approximately universal code design criterion \cite{TW05} and the non-vanishing determinant criterion \cite{BelRek03,YaoWor03} results in DM-tradeoff optimal codes. Finally, we note that the concepts introduced in this paper can be extended to multiple-access selective-fading MIMO channels and to the analysis of the DM-tradeoff properties of specific (suboptimal) receivers.
\bibliographystyle{IEEEtran}
\bibliography{biblio}

\end{document}